\documentclass[aps, 10pt,reprint,superscriptaddress,showpacs,longbibliography,floatfix]{revtex4-2}

\usepackage{xcolor}
\usepackage{physics}
\usepackage{amssymb}
\usepackage{graphicx}
\usepackage{mathtools}
\usepackage{amsmath}
\usepackage{hyperref}
\usepackage{tikz}
\usepackage{amsfonts}
\usepackage{tcolorbox}
\usepackage{braket}
\hypersetup{
colorlinks = true,
urlcolor = blue,
linkcolor = blue,
citecolor = blue
}

\newcommand{\TUM}{\affiliation{Technical University of Munich, TUM School of Natural Sciences, Physics Department, 85748 Garching, Germany}}
\newcommand{\MCQST}{\affiliation{Munich Center for Quantum Science and Technology (MCQST), Schellingstr. 4, 80799 M{\"u}nchen, Germany}}
\newcommand{\MPIPKS}{\affiliation{Max Planck Institute for the Physics of Complex Systems, N{\"o}thnitzer Str. 38, 01187 Dresden, Germany}}

\begin{document}
\author{Melissa Will} \TUM \MCQST
\author{Roderich Moessner}\MPIPKS
\author{Frank Pollmann} \TUM \MCQST

\title{Realization of Hilbert Space Fragmentation and Fracton Dynamics in 2D}

\begin{abstract}
We propose the strongly tilted Bose-Hubbard model as a natural platform to explore Hilbert-space fragmentation (HSF) and fracton dynamics in two-dimensions, in a setup and regime readily accessible in optical lattice experiments. Using a perturbative ansatz, we find HSF when the model is tuned to the resonant limit of on-site interaction and tilted potential. First, we investigate the quench dynamics of this system and observe numerically that the relaxation dynamics strongly depends on the chosen initial state—one of the key signature of HSF. Second, we identify fractonic excitations with restricted mobility leading to anomalous transport properties. Specifically, we find excitations that show one dimensional diffusion ($z=1/2$) as well as excitations that show subdiffusive behaviour in two dimensions ($z=3/4$). Using a cellular automaton, we analyze their dynamics and compare to an effective hydrodynamic description.

\end{abstract}
\maketitle
An important question in the context of quantum-many-body systems out of equilibrium is how thermodynamics arises in closed systems. 
Generic quantum systems are expected to reach a thermal equilibrium as predicted by the Eigenstate Thermalization Hypothesis (ETH) \cite{deutsch1991quantum,rigol2008thermalization,srednicki1994chaos,d2016quantum}, where the thermal properties of the thermal equilibrium state are determined by the conserved quantities (such as energy or particle number). 
Several mechanisms challenging this concept have been proposed recently.
Notably, many-body localization (MBL) \cite{basko2006metal,nandkishore2015many,abanin2019colloquium,altman2015universal,schreiber2015observation} is a candidate for breaking ergodicity and defy thermalization for all eigenstates in the presence of sufficiently strong disorder.
A weaker form of ergodicity breaking occurs in systems with quantum many-body scar states, which are atypical non-thermal states in an otherwise thermal spectrum 
\cite{turner2018weak,moudgalya2022quantum,moudgalya2018exact,Moudgalya_2018}.

Systems with constrained dynamics have been shown to exhibit very rich non-equilibrium dynamics and to provide novel mechanisms to break ergodicity.
Constrained dynamics occurs, for example, naturally in systems with dipole conservation, where the dynamics always involves a collective motion of several particles (e.g., squeezing dynamics). 
In a series of recent works \cite{sala2020ergodicity,khemani2020localization,moudgalya2022quantum}, it has been shown that one-dimensional (1D) systems with constrained dynamics can exhibit \emph{Hilbert-space fragmentation} (HSF). 
In this scenario, the Hilbert space splits into exponentially many distinct sectors in some simple local basis, which are dynamically disconnected from one another (i.e., Krylov sectors). 
This is in stark contrast to ordinary global symmetries, which lead to a polynomial number of sectors.
In the case of \emph{strong fragmentation}, the size of the largest connected sector is exponentially smaller than the total number of states, leading to a complete breakdown of ETH and true localization. 
In the case of \emph{weak fragmentation}, the constrained dynamics is no longer sufficient to strongly violate ETH, i.e., most eigenstates are thermal. 
\begin{figure}
    \centering
    \tikz\path
  node[inner sep=0pt] (system) 
    {\includegraphics[width=1\linewidth]{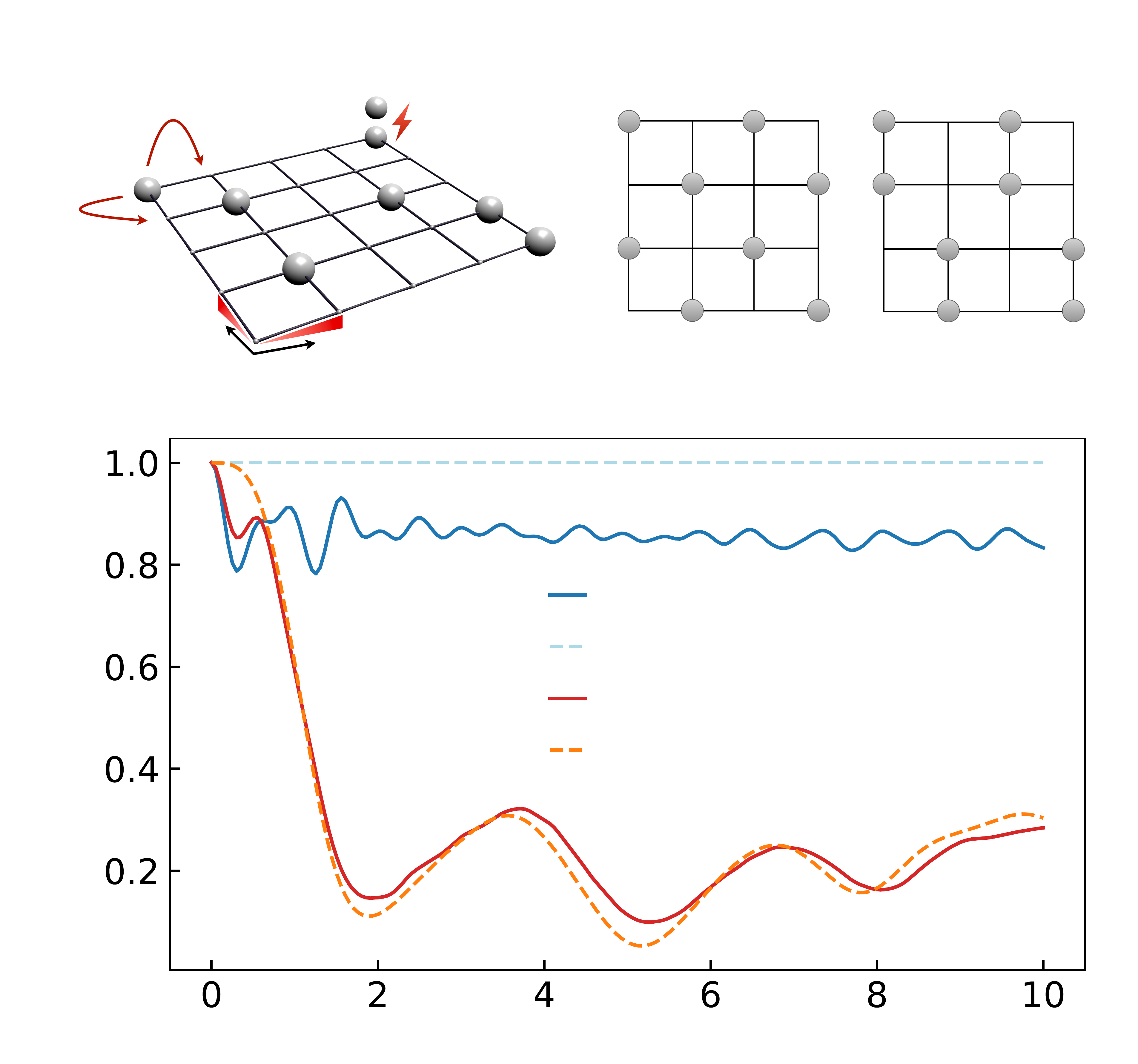}}
    (-4,3.4) node {\textbf{a}}
    (-3.4,2.9) node {$J$}
    (-1,3.1) node {$U$}
    (-2.85,1.8) node {$\Delta$}
    (-1.55,1.55) node {$\Delta$}
    (-2,1.25) node {$x$}
    (-2.65,1.35) node {$y$}
    (0.2,3.4) node {\textbf{b}}
    (2.2,3.4) node {\textbf{c}}
    (-4,1.) node {\textbf{d}}
    (-4,-1.2) node [rotate=90] {$\mathcal{I}$}
    (0.3,-3.8) node {$t$}
    (1.65,-0.45) node {Checkerboard exact}
    (1.85,-0.85) node {Checkerboard effective}
    (1.15,-1.25) node {Dimer exact}
    (1.35,-1.65) node {Dimer effective}
    (3,1) node {$\Delta = U \gg J$}
  ;
    \caption{\textbf{Illustration of the setup and dynamics of the imbalance in different Krylov sectors.} \textbf{a} Schematic of the tilted, two dimensional Bose-Hubbard model with the tilted potential of strength $\Delta$, tunneling $J$ and Hubbard interaction $U$. In the limit $\Delta=U\gg J$ an effective Hamiltonian features Hilbert-space fragmentation. \textbf{b} The checkerboard state is completely frozen under effective time evolution. \textbf{c} In contrast the dimer state is highly mobile and part of the largest Krylov sector. \textbf{d} Imbalance shows strong dependence on the initial state regarding its dynamics. Parameters: $L=4,J=1,U=10,\Delta=10$.}
    \label{fig:Figure1}
\end{figure}
Autocorrelators of local operators, for example, decay to their thermal values for typical states, albeit in certain cases exhibiting subdiffusive transport \cite{feldmeier2020anomalous,Gromov2020}.
However, such systems still violate the strong version of ETH and exhibit exponentially many non-thermal eigenstates. 
Remarkably, several hallmarks of HSF were observed experimentally in a tilted 1D Fermi-Hubbard model \cite{kohlert2021experimental,scherg2021observing}. 
Beyond 1D, a number of higher dimensional models have been proposed theoretically that exhibit HSF in higher dimensions \cite{lehmann2023fragmentation,sala2022dynamics,khudorozhkov2022hilbert}. 

In this paper, we propose a Bose-Hubbard model as a natural platform to observe HSF in a two-dimensional (2D) lattice system and will discuss in the end, that the model and the required parameter regime are readily accessible in optical lattice experiments.
We demonstrate that HSF fragmentation occurs in the limit of a strongly tilted potential when the repulsive onsite interaction is tuned to resonance.
Specifically, we show that the experimentally accessible imbalance shows a strong initial state dependent dynamics for states with the same quantum numbers and energy, which is a key signature of HSF. 
The emergent HSF is understood by perturbatively deriving an effective model for which the fragmentation can be explicitly shown. 
In addition, we find that the system hosts fractonic \cite{Haah2011} excitations with restricted mobility, leading to anomalous transport properties. We identify two distinct excitations, where one diffuses ordinarily on a one-dimensional strip and the other in two-dimensions showing subdiffusion in one direction and ordinary diffusion in the other.

\paragraph*{\textbf{Bose-Hubbard model.}}
We consider the Bose-Hubbard model on a square lattice in the presence of a tilted onsite potential, described by the Hamiltonian
\begin{equation}\label{eq:Hamiltonian}
    \begin{split}
    \mathcal{H} &= -J \sum_{\langle (x,y),(x',y')\rangle} \left(b^\dagger_{(x,y)}b_{(x',y')} +h.c. \right)\\
    &+\sum_{(x,y)}\left(\Delta(x+y)n_{(x,y)} +\frac{U}{2}n_{(x,y)}\left(n_{(x,y)}-1\right)\right), \end{split} 
\end{equation}
where the operator $b^{(\dagger)}_{(x,y)}$ annihilates (creates) a boson on site \mbox{$x,y$}. 
The tunneling and repulsive onsite interaction strength are denoted by $J$ and $U$, respectively. 
We tune the onsite interaction into resonance with the tilted potential and take it to be the largest energy scale, i.e. $U=\Delta\gg J$. 
In this limit, a boson without any occupied nearest-neighbor sites cannot move due to a large energy penalty, whereas correlated hopping involving a boson on a nearest-neighbor site is possible due to the resonance condition (details will be discussed below). 
A configuration with one boson on each site of one of the two sublattices (checkerboard pattern) is thus frozen as none of the bosons have any occupied neighbouring sites, whereas having a pairwise pattern (dimer state) leads to a highly mobile configuration (see Fig.~\ref{fig:Figure1}b-c). 
This state dependent dynamics is reflected by the imbalance shown in Fig.~\ref{fig:Figure1}d. 
The imbalance is an experimentally accessible quantity comparing the filling of initially occupied sites to later times \cite{scherg2021observing}, given by:
\begin{align}
    \mathcal{I} = \frac{\sum_{(x,y)\in A}n_{x,y}-\sum_{(x,y)\not\in A}n_{x,y}}{N_{\text{total}}},
\end{align}
where $A$ contains all sites which are initially occupied and $n_{x,y}$ is the number of particles on site $(x,y)$. 
A non-zero steady-state imbalance signals a memory of the initial state for $\mathcal{I}(t=0)=1$ and thus implies a non-thermal state. 
For $U=\Delta =10J$, we observe that the imbalance of the checkerboard state stays close to one, whereas for the dimer configuration decreases quickly. 
Notably, both configurations have the same quantum numbers (i.e., particle number and energy). 
This behavior is reminiscent of systems with quantum many-body scars, which exhibit an initial state dependent dynamics \cite{turner2018weak,moudgalya2022quantum}. 
In the following, we  first explain this phenomenon by deriving an effective Hamiltonian, which features HSF and yields fractonic excitations. 
\begin{figure}
    \centering
    \tikz\path
  node[inner sep=0pt] (system) 
    {\includegraphics[width=1\linewidth]{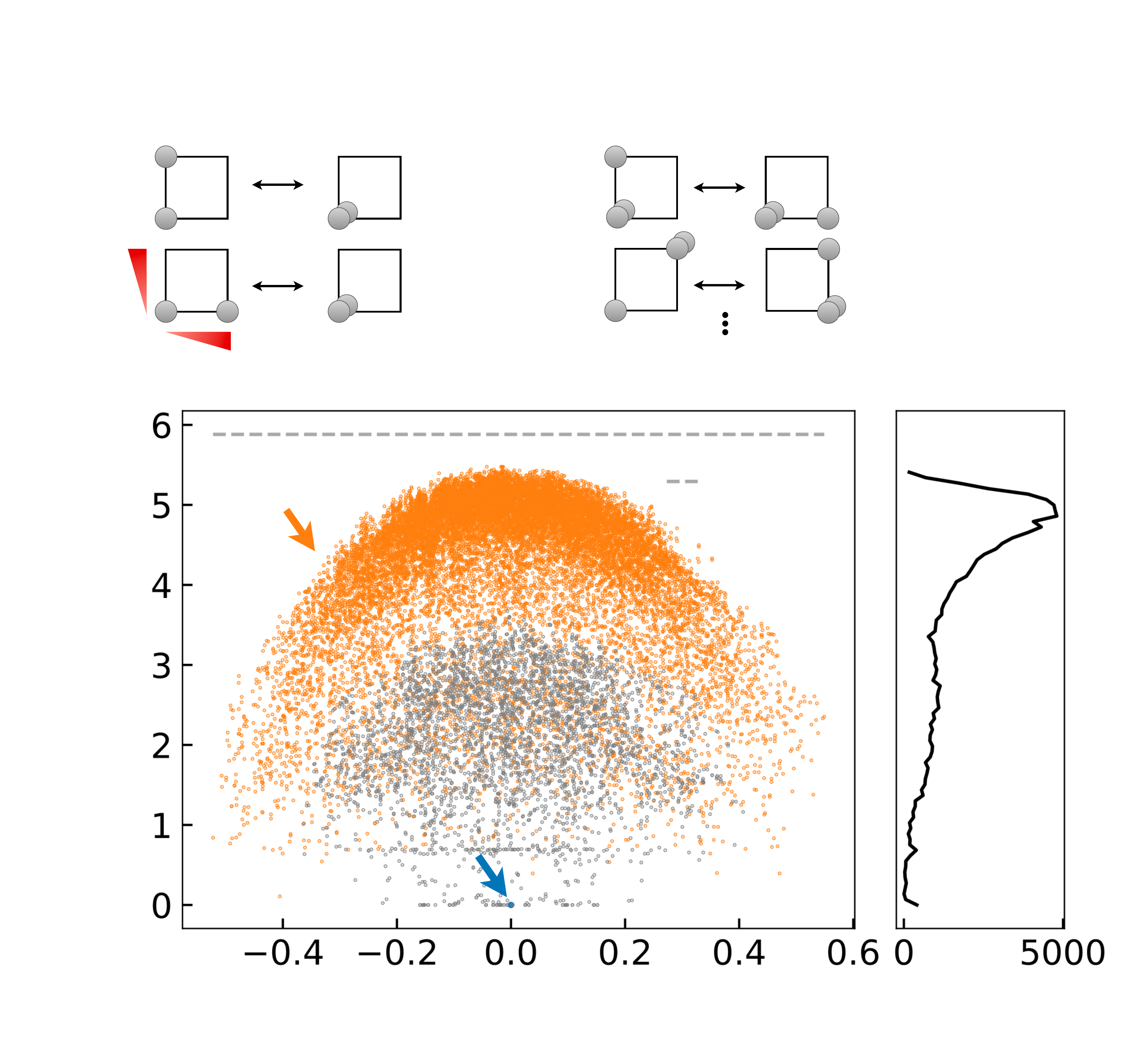}}
    (-3,3.2) node {\textbf{a} $1^{\text{st}}$ oder}
    (0.8,3.2) node {\textbf{b} $2^{\text{nd}}$ order}
    (-1,2.5) node {$J$}
    (-1,1.8) node {$J$}
    (-3.3,2.2) node {$\Delta$}
    (-2.4,1.3) node {$\Delta$}
    (2.5,2.5) node {$\frac{J^2}{\Delta}$}
    (3,1.8) node {$\frac{-J^2}{\Delta+U}+\frac{J^2}{\Delta}$}
    (-3.6,1.) node {\textbf{c}}
    (-3.6,-1.2) node [rotate=90] {$\mathcal{S}$}
    (-0.45,-3.6) node {$E/N$}
    (3.1,-3.6) node {$\#$}
    (1.4,0.3) node {$S_{\text{Page}}$}
    (-1.9,0.3) node {Dimer sector}
    (-1.6,-2.6) node {Chk sector}
  ;
    \caption{\textbf{Effective Hamiltonian and entanglement of its eigenstates.} \textbf{a} Processes with one hop, that leave $E_{\Delta+U}$ of a configuration unchanged. \textbf{b} Examples of second order processes obtained with degenerate perturbation theory and their matrix elements are shown. A complete list with all second order processes can be found in App.~\ref{sec:appendix_perturbation_theory}, see Fig.~\ref{fig:perturbation_theory}.~\textbf{c} Entanglement entropies of eigenstates for a bipartition into two halves for all sectors with energy \mbox{$E_{\Delta+U}=E_{\text{chk}}$}. The Krylov sector containing the dimer state is coloured in orange and the checkerboard state in blue. The histogram on the right shows that most of the states lie close to the Page value. Parameters: $L=4,J=1,U=10,\Delta=10$.}
    \label{fig:Figure2}
\end{figure}

\paragraph*{\textbf{Effective model.} }
To obtain an effective description, we employ degenerate perturbation theory \cite{soliverez1969effective,mila2010strong} up to second order. 
The tilt $\Delta$ and interaction strength $U$ set the system's largest energy scale, and the hopping term $J$ is assumed to be small. 
Thus all terms in the effective description conserve the energy
\begin{equation}\label{eq:equipotential}
    \begin{split}
     E_{\Delta +U}&= \sum_{(x,y)} \left( \vphantom{\frac{U}{2}} \Delta(x+y)n_{(x,y)} \right.\\
     & \left. + \frac{U}{2}n_{(x,y)}\left(n_{(x,y)}-1 \right) \right).
    \end{split}    
\end{equation}
Figure~\ref{fig:Figure2}a shows all first order processes and Fig.~\ref{fig:Figure2}b shows selected second order processes. 
For example, doublons can resonantly form out of, and decay into, a pair of neighbouring bosons in first order. 
A complete list of all hopping terms and further details can be found in App.~\ref{sec:appendix_perturbation_theory}. 
Note that all hopping processes require bosons on neighboring sites, leading to strong constraints on the dynamics. 
Figure~\ref{fig:Figure1}d shows that under time evolution with the effective Hamiltonian, the checkerboard state is completely frozen, and the dimer configuration is highly mobile. 
Moreover, we find good quantitative agreement with the previously discussed dynamics under time evolution with the Bose-Hubbard model. 
Thus, the state dependent dynamics originate from the constrained dynamics of the effective Hamiltonian. 
In particular, we find that the dimer configuration is part of the largest Krylov sector, whereas the checkerboard state is a frozen state that is not connected to any other state. 

While we can numerically only access small clusters, we find several hallmarks of \emph{weak HSF}:
(i) The number of Krylov sectors as well as the number of frozen states grows exponentially with system size, showing that the system features HSF. 
A detailed derivation of this scaling can be found in App.~\ref{sec:Scaling_Kryl_sec}. 
(ii) The bipartite entanglement entropy of most eigenstates of the effective Hamiltonian approaches the Page value \cite{page1993average} for most eigenstates, while only a small subset is only slightly entangled, see Fig.~\ref{fig:Figure2}c, where we consider a vertical bipartition into two halves of the system. 
(iii) The frozen site density \cite{morningstar2020kinetically,sala2020ergodicity} vanishes, which can, therefore not be used as a positive indication for strong fragmentation. 
Moreover, the largest connected sector spans a sizable fraction of the Hilbert space at $E_{\Delta+U}=E_{\text{chk}}=E_{\text{dim}}$ (\mbox{$r = 83\%$} for \mbox{$L\times L=4\times 4$} and $r= 66\%$ for \mbox{$L \times L = 5\times 5$}.
Taking those findings together, they hint towards weak HSF.\\

\paragraph*{\textbf{Fracton Dynamics.}} 
\label{sec:Defects}
\begin{figure}
    \centering
    \tikz\path
  node[inner sep=0pt] (system) 
    {\includegraphics[width=1\linewidth]{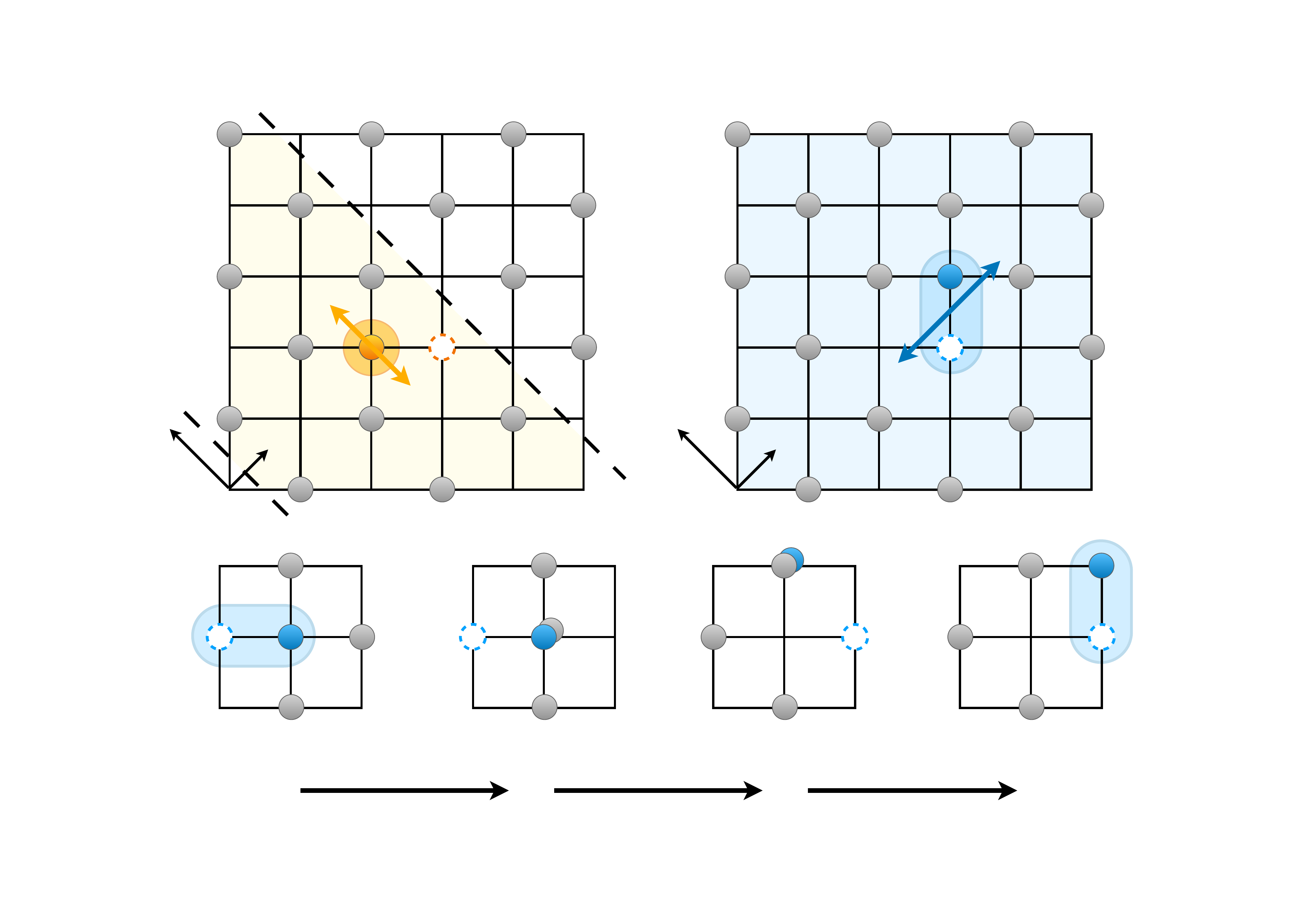}}
    (-2.2,2.5) node {\textbf{a} `Negative' defect}
    (1.3,2.5) node {\textbf{b} `Positive' defect}
    (-3.3,-0.8) node {\textbf{c}}
    (-3.25,0) node [rotate=45] {$\tilde{y}$}
    (0.05,-0) node [rotate=45] {$\tilde{y}$}
    (0.84,0.08) node [rotate=45] {$\tilde{x}$}
    (-2.49,0.08) node [rotate=45] {$\tilde{x}$}
    (-1.65,-2.4) node {$1^{\text{st}}$ord.}
    (1.65,-2.4) node {$1^{\text{st}}$ord.}
    (0,-2.4) node {$2^{\text{nd}}$ord.}
  ;
    \caption{\textbf{Defects in the checkerboard state with restricted mobility.} \textbf{a} Displacing one boson of the checkerboard state to the left or down, decreases the states energy and creates a `negative' defect (in orange) and leaves a hole (white sphere). Such a `negative' defect is restricted to hopping on a stripe in $\tilde{y}-$direction (shaded in yellow, between dashed lines). The hole cannot move along with the defect. \textbf{b} In contrast, increasing one boson's energy, by moving it to the right or up, creates a `positive' defect (blue sphere), which can move in the entire 2D plane (shaded in blue). Movement in $\tilde{x}-$direction happens together with its hole due to energy conservation. In $\tilde{y}-$direction the defects motion is decoupled from its hole. \textbf{e} One exemplary process for the `positive' defect and its hole to travel in $\tilde{x}-$direction consist out of two hopping processes contained in $\mathcal{H}^{(1)}$ and one second order process, see Fig.~\ref{fig:Figure2}.}
    \label{fig:Figure3}
\end{figure}
\begin{figure}
    \centering
    \tikz\path
  node[inner sep=0pt] (system) 
    {\includegraphics[width=1\linewidth]{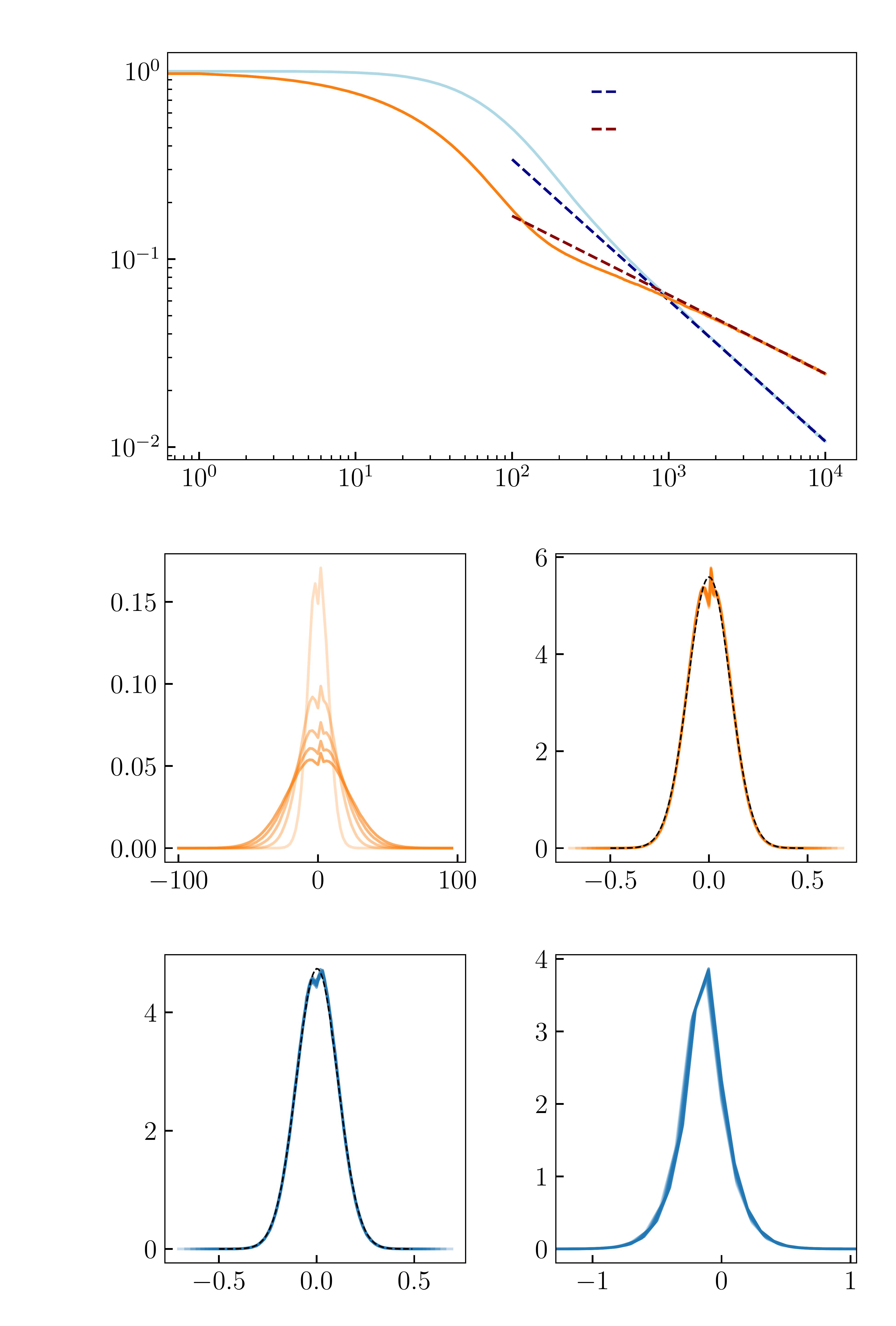}}
    (-3.4,6.1) node {\textbf{a}}
    (-3.4,1.2) node {\textbf{b}}
    (0.6,1.2) node {\textbf{c}}
    (-3.4,-2.7) node {\textbf{d}}
    (0.6,-2.7) node {\textbf{e}}
    (-3.4,3.9) node [rotate=90] {$\mathcal{C}$}
    (2.7,5.55) node {`positive' defect}
    (2.75,5.15) node {`negative' defect}
    (0.55,1.5) node {$t$}
    (3.3,0.7) node {$z = 0.5$}
    (3.3,-3.1) node {$z = 0.25$}
    (-0.5,-3.1) node {$z = 0.5$}
    (-0.4,0.7) node {$t\lbrack 10^3\rbrack:$}
    (-0.3,0.35) node {$1$}
    (-0.3,0) node {$3$}
    (-0.3,-0.4) node {$5$}
    (-0.3,-0.75) node {$8$}
    (-0.2,-1.1) node {$10$}
    (-1.3,-2.4) node {$\tilde{y}$}
    (2.65,-2.4) node {$\tilde{y}t^{-z}$}
    (2.65,-6.3) node {$\tilde{x}t^{-z}$}
    (-1.3,-6.3) node {$\tilde{y}t^{-z}$}
    (0.6,-4.3) node [rotate=90]{$ \langle n(\tilde{x},t) \rangle t^{z}$}
    (-3.2,-4.3) node [rotate=90]{$ \langle n(\tilde{y},t) \rangle t^{z}$}
    (0.6,-0.5) node [rotate=90]{$ \langle n(\tilde{y},t) \rangle t^{z}$}
    (-3.5,-0.5) node [rotate=90]{$ \langle n(\tilde{y},t) \rangle$}
    (3.3,3.5) node {$\sim t^{-0.42}$}  
    (2.5,2.5) node {$ t^{-0.75}\sim$}  
;
    \caption{\textbf{Diffusion of defects.} \textbf{a} Return probability $\mathcal{C}(0,0,t)$ for the `positive' and `negative' defect obtained from a cellular automaton evolution. The long time behavior approaches an algebraic decay $\sim t^{-z}$. The numerical values of the exponent $z$ were extracted from fits starting at $t=5\times 10^3$ (dashed lines). \textbf{b} Density profile in $\tilde{y}-$direction of one `negative' defect initially in the middle of the system at different time steps. The $\tilde{x}-$direction has been integrated over. \textbf{c} Scaling collapse for `negative' defect with critical exponent of $z=0.5$ shows ordinary 1D diffusion. \textbf{d-e} Scaling collapse for `positive' defect in $\tilde{y}-$ and $\tilde{x}-$direction. Critical exponents show ordinary diffusion in $\tilde{y}-$direction and sub\-diffusion in $\tilde{x}-$direction. A Gaussian profile is shown in dashed for diffusion in $\tilde y-$direction. Parameters: $L=100, N_{\text{av}}=10^5$.}
    \label{fig:Figure4}
\end{figure}

We will now investigate excitations on top of the frozen checkerboard state. 
We find two different types of defects, as shown in Fig~\ref{fig:Figure3}\mbox{a-b}: A single defect can be formed by either decreasing the energy of a boson, e.g., moving it one site to the left or down (`negative' defect) or creating a `positive' defect by increasing its energy. 
The `negative' defect can only move within a 1D strip, whereas a `positive' defect can move in the entire 2D system, respectively. Such excitations with restricted mobility are known as fractons and have been found in constrained quantum systems \cite{pretko2020fracton}.

We begin by discussing the mobility and emergent hydrodynamics of the `negative' defect and then turn to the `positive' one. 
The mobility of a `negative' defect is determined by the hopping terms of the effective Hamiltonian, which all conserve the energy $E_{\Delta+U}$, see Eq.~\eqref{eq:equipotential}.
Therefore, a defect can either move on its own on an equipotential of $E_{\Delta+U}$ or conserve $E_{\Delta+U}$ by moving together with an adjacent hole. 
Here, the hole refers to the vacant site within the checkerboard pattern, shown as a dashed circle in Fig.~\ref{fig:Figure3}a. 
The equipotential of $E_{\Delta+U}$ contains quasi-1D lines in $\tilde y-$direction, as indicated by the yellow marking in Fig.~\ref{fig:Figure3}a. 
There exists no hopping term in the effective Hamiltonian which moves the `negative' defect together with its hole away from its initial equipotential line.
Thus the `negative' defect is restricted to this quasi-1D line in $\tilde{y}-$direction.
Since the broadening of the quasi-1D line in $\tilde{x}-$direction can be neglected in the long wavelength limit, the diffusion of a single defect can be understood using hydrodynamics in 1D. 
The only conserved quantity in $\tilde{y}-$direction is particle number, leading to an ordinary 1D hydrodynamic equation given by
\begin{align}
\partial_t n^-(\tilde{y},t) = -\partial^2_{\tilde{y}}n^-(\tilde{y},t)
\label{eq:dif_y}
\end{align}
with $n^-(\tilde{y},t)$ number of particles on site $\tilde{y}$ at time $t$.
We therefore expect, that the onsite-correlation function
\begin{align}
C_{\tilde{x},\tilde{y}}(\left(t \right) = \langle n(\tilde{x},\tilde{y},t)n(0,0,0)\rangle
\end{align}
with $\langle...\rangle$ denoting an average over different trajectories, scales as \mbox{$C_{0,0}^-(t)\sim t^{-0.5}$}. In the following, we  refer to the absolute value of the exponent as $z$.

To study the late-time dynamics numerically, we employ a classical cellular automaton, which mimics the constrained dynamics of the Hamiltonian \cite{feldmeier2020anomalous,ritort2003glassy}. 
The discrete-time evolution consists of local updates, which are given by the hopping terms of the effective Hamiltonian. 
Thus they respect the same conservation laws and map product states to product states. For further details, see App.~\ref{sec:Cellular_Automaton}. 
Figure~\ref{fig:Figure4}a shows the onsite correlation function over time and a fit extrapolating the late-time decay exponent. For the `negative' defect, we estimate a scaling of \mbox{$C_{0,0}^-(t)\sim t^{-0.42}$}. 
Figure~\ref{fig:Figure4}b shows the density profile of the `negative' defect in $\tilde{y}-$direction for several time steps while the $\tilde{x}-$direction is integrated out. 
Performing a scaling collapse, we find that those curves collapse for a late-time decay exponent of $z^-=0.5$, shown in Fig.~\ref{fig:Figure4}c. 
Furthermore, the profile is close to a Gaussian, shown in dashed black. 
This result reflects that the `negative' defect can only move on a quasi-1D line.

In contrast, we find that a `positive' defect is mobile in the entire 2D system, see Fig.~\ref{fig:Figure3}b. 
Additionally to the movement on an equipotential line, the effective Hamiltonian contains hopping processes which move the `positive' defect together with its hole in $\tilde{x}-$direction. 
Figure~\ref{fig:Figure3}c shows an example of one combination of such processes. At first, a doublon is formed via a first-order process. 
Then a second order process moves the doublon one site up, and it finally decays via a first-order process. 
Ultimately, the `positive' defect has traveled together with its hole along $\tilde{x}-$direction. 
At late times, the hydrodynamics of the `positive' defect in the $\tilde{y}-$direction is also described by Eq.~(\ref{eq:dif_y}). To obtain the diffusion equation in the $\tilde{x}-$direction, we analyze Fig.~\ref{fig:Figure3}c and find that the quantities conserved at the beginning and at the end of the process are particle number and dipole moment. 
References \cite{feldmeier2020anomalous,Gromov2020} showed that for a 1D system with particle and dipole moment conservation, the diffusion equation is given by \mbox{$\partial_t n(x,t) = \partial^4_{x} n(x,t)$} and the onsite-correlation function shows a subdiffusive scaling of $C(t)\sim t^{-0.25}$.
Using this result, the diffusion equation of the `positive' defect is in total given by:
\begin{align}
\label{eq:pos_hydro}
\partial_t n^+(\tilde{x},\tilde{y},t) = -\partial^2_{\tilde{y}}n^+(\tilde{x},\tilde{y},t)+\partial^4_{\tilde{x}}n^+(\tilde{x},\tilde{y},t).
\end{align}
This combines to an overall 2D subdiffusive late-time scaling exponent of \mbox{$z^+ = z^+_{\tilde{x}}+z^+_{\tilde{y}} = 0.25+0.5=0.75$}, whereas ordinary diffusion in 2D scales with $z_{2\text{D}}=1$. 
Figure~\ref{fig:Figure4}a shows that the onsite-correlation function indeed decays with $C^+\sim t^{-0.75}$. 
Figure~\ref{fig:Figure4}d-e show scaling collapses for the `positive' defect in $\tilde{y}-$ and $\tilde{x}-$direction. 
We find that the `positive' defect ordinarily diffuses in $\tilde{y}-$direction with $z^+_{\tilde{y}}=0.5$ whereas sub-diffusively in $\tilde{x}-$direction with an exponent of $z^+_{\tilde{x}}=0.25$. 

\paragraph*{\textbf{Conclusions.}}
In this work, we have demonstrated the emergence of HSF and fractonic excitations in a strongly tilted Bose-Hubbard model, when the repulsive onsite interaction is tuned into resonance with the tilt. 
The proposed system is shown to exhibit an initial state dependant dynamics as a signature of HSF.
Moreover, we identified fractonic excitations with highly anisotropic, anomalous diffusion.
We emphasize that the model and the required parameter regime are readily accessible in optical lattice experiments \cite{gross2017quantum}. 
State of the art experiments also allow to create a wide range of initial states and to track the long time dynamics using single atom resolution quantum gas microscopes \cite{bakr2009quantum,sherson2010single}. 
Thus, we expect that the model will allow to observe clear signatures of 2D HSF in readily existing optical lattice experiments. 
As for the fracton dynamics, we focused here only on single excitations and conjecture that the signatures should be robust at sufficiently small densities but leave the detailed study for a future work.
Lastly, while we investigated a 2D model, the resonance condition between a strong tilt and on-site interaction similarly leads to intricate constrained dynamics and likely HSF also in 3D models.

\begin{acknowledgements}
The authors thank Pablo Sala, Johannes Feldmeier, Alexey Khudorozhkov, Johannes Zeiher and his group, Monika Aidelsburger, Immanuel Bloch and Fabian Heidrich-Meisner for helpful discussions. This research was financially supported by the European Research Council (ERC) under the European Union’s Horizon 2020 research and innovation program under grant agreement No.~771537. F.P. acknowledges the support of the Deutsche Forschungsgemeinschaft (DFG, German Research Foundation) under Germany’s Excellence Strategy EXC-2111-390814868 and DFG Research Unit FOR 5522 (project-id 499180199). F.P.’s research is part of the Munich Quantum Valley, which is supported by the Bavarian state government with funds from the Hightech Agenda Bayern Plus. This work was in part supported by the Deutsche Forschungsgemeinschaft under grants SFB 1143 (project-id 247310070) and the cluster of excellence ct.qmat (EXC 2147, project-id 390858490).\end{acknowledgements}
\paragraph*{\textbf{Data and materials availability.}} Data analysis and simulation codes are available on Zenodo upon reasonable request \cite{zenodo}.

\bibliography{Bib.bib}

\appendix
\section{Perturbation theory}
\label{sec:appendix_perturbation_theory}
\begin{figure*}
    \centering
    \includegraphics[width=0.85\linewidth]{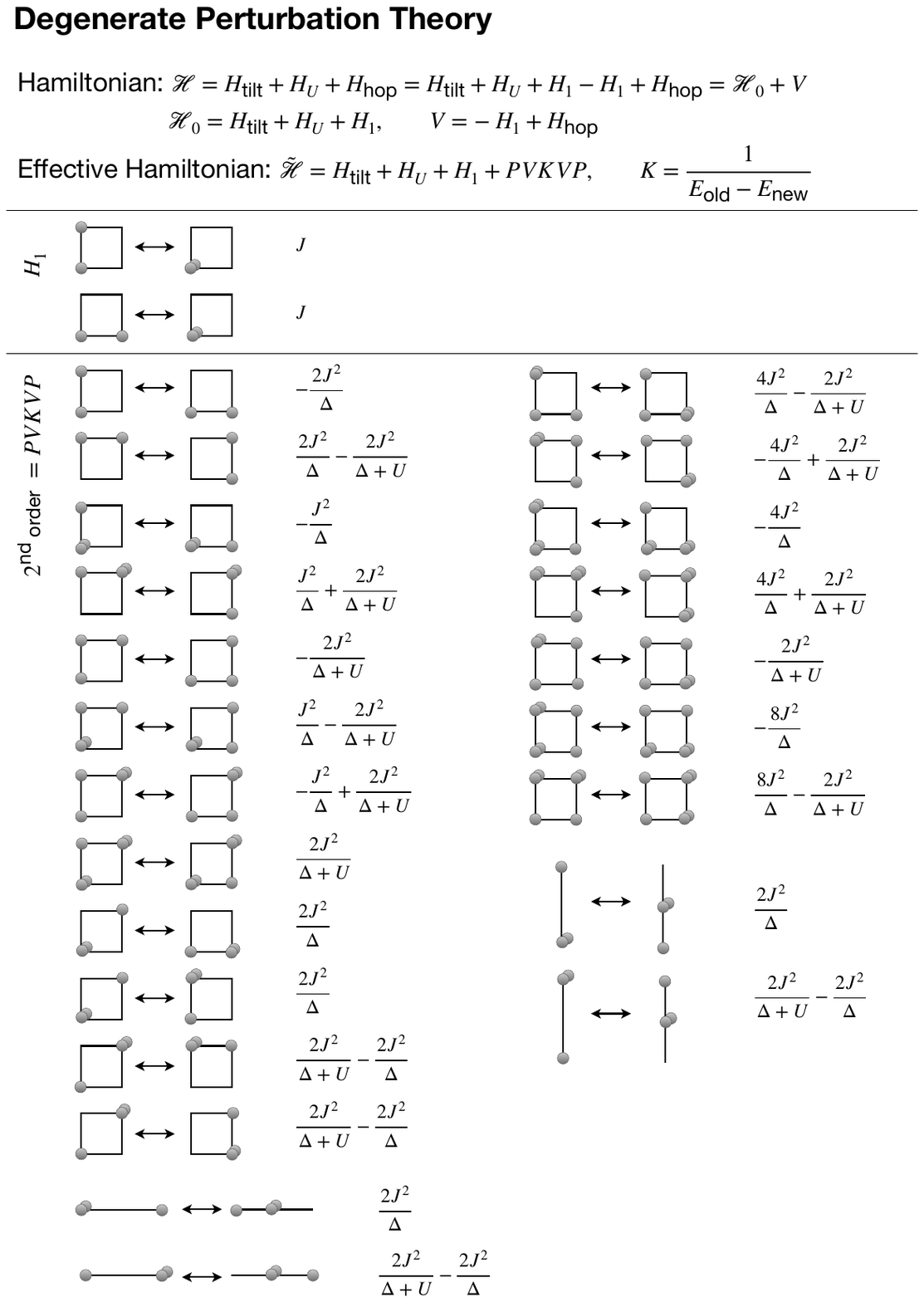}
    \caption{\textbf{Perturbation theory} All relevant processes up to second order for the checkerboard and dimer state are shown. }
    \label{fig:perturbation_theory}
\end{figure*}

To obtain an effective description, we employ degenerate perturbation theory. The tilt $\Delta$ and interaction strength $U$ set the system's largest energy scale, and the hopping term gives the perturbation. Figure~\ref{fig:Figure2}a shows hopping processes, which do not change the state's energy with respect to tilt and energy. They are given by:
\begin{align}
    \mathcal{H}^{(1)} &= -P\mathcal{H}_{\text{hop}}P
\end{align}
with $P$ being the projector into an energy subspace of \mbox{$\mathcal{H}_{\text{tilt}} + \mathcal{H}_U$}. Those hopping terms $\mathcal{H}^{(1)}$ have to be excluded from the perturbation $V$. We define:
\begin{align}
    \mathcal{H} &= \mathcal{H}_0 +V \\
    &= \mathcal{H}_{\text{tilt}} + \mathcal{H}_U + \mathcal{H}^{(1)}-\mathcal{H}^{(1)}+\mathcal{H}_{\text{hop}} \\
    \mathcal{H}_0 &=\mathcal{H}_{\text{tilt}} + \mathcal{H}_U + \mathcal{H}^{(1)} \label{eq:H0}\\
    V &= \mathcal{H}_{\text{hop}}-\mathcal{H}^{(1)} \label{eq:V}.
\end{align}
The effective Hamiltonian up to second order is then given by:
\begin{align}
   \tilde{\mathcal{H}} &= \mathcal{H}_0 + \mathcal{H}^{(2)}\\
    \mathcal{H}^{(2)} &= PVKVP
\label{eq:degenerate_perturbation_theory}
\end{align}
with \mbox{$K = 1/\left(E_{\text{new}}-E_{\text{old}}\right)$}. A list of all processes and energies, which are relevant for the checkerboard and dimer state are listed in Fig.~\ref{fig:perturbation_theory}.

\section{Scaling of number of Krylov sectors}
\label{sec:Scaling_Kryl_sec}
\begin{figure}
    \centering
    \centering
    \tikz\path
    node[inner sep=0pt] (frozenstates) 
    {\includegraphics[width=1\linewidth]{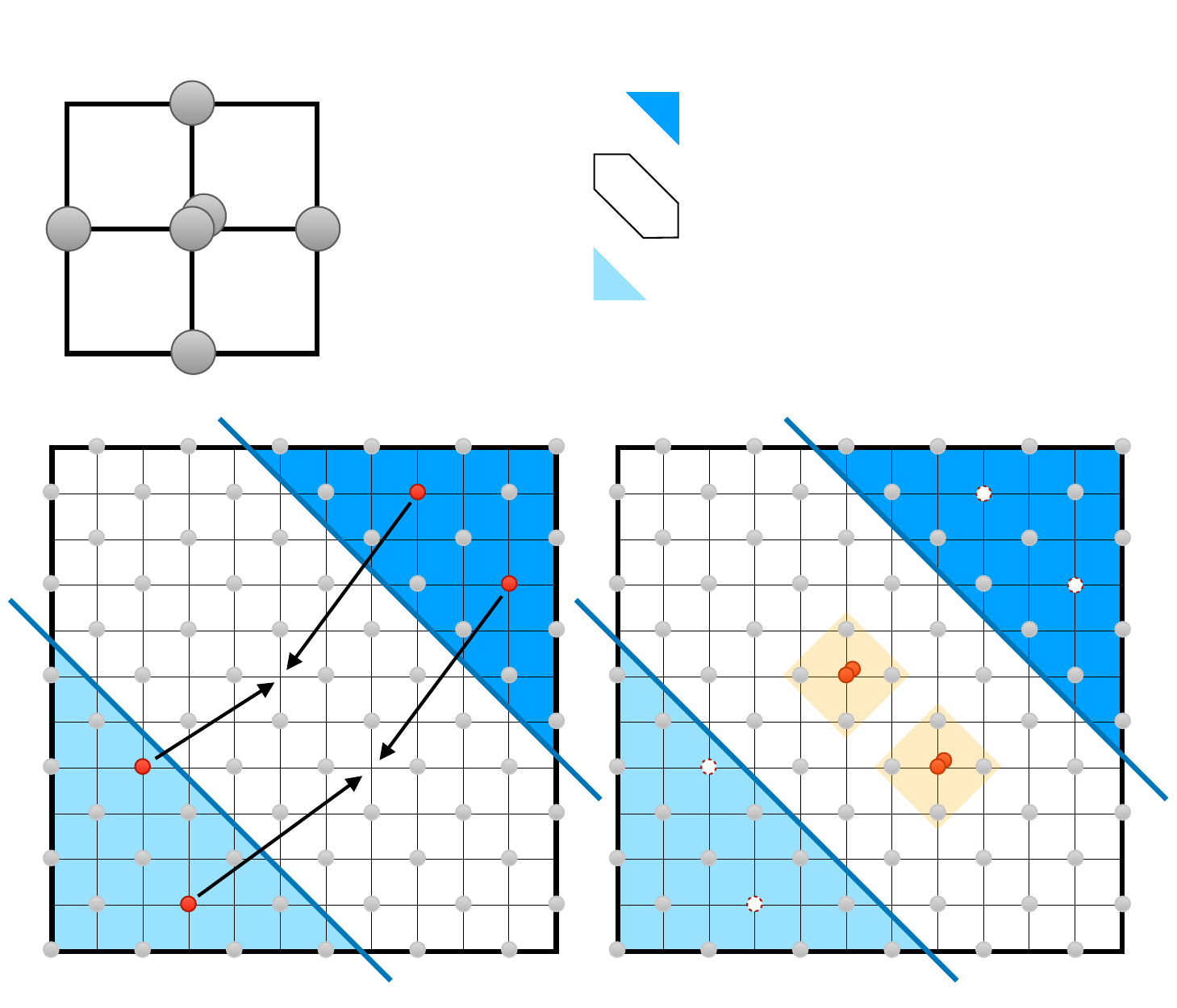}}
    (-2.8,3.4) node {\textbf{a} Frozen pattern}
    (1,3.4) node {Area of region:}
    (1.2,2.9) node {$\sim \frac{L^2}{3}$}
    (1.2,2.3) node {$\sim \frac{L^2}{3}$}
    (1.2,1.7) node {$\sim \frac{L^2}{3}$}
    (-3.9,0.7) node {\textbf{b}}
    (0.3,0.7) node {\textbf{c}};
    \caption{\textbf{Creation of frozen states.} \textbf{a} Frozen pattern of particles, which cannot be rearranged by the effective Hamiltonian. \textbf{b} In order to create a frozen configuration containing such a frozen pattern, one can choose an equal number of particles from the lower left area and respectively from the upper right one. Moving them to the sites indicated by arrows, generates the frozen pattern with same $E_{\Delta+U}$. \textbf{c} New configuration with two frozen patterns in the middle region. }
    \label{fig:Krylovsectorscaling}
\end{figure} 
In the following, we will first show that the number of frozen states scales exponentially with system size and will then generalize the argument to show that also the number of mobile sectors grows exponentially with system size. For this, we identify the pattern shown in Fig.~\ref{fig:Krylovsectorscaling}a as frozen and derive a lower bound by estimating the number of possibilities to create such patterns. Starting from the checkerboard state, we divide the system into three approximately equal-sized parts, see Fig.~\ref{fig:Krylovsectorscaling}b. The lower left (light blue) and upper right (dark blue) corners each contain approximately $\frac{L^2}{3}$ sites and, therefore $\frac{L^2}{6}$ particles. The idea is to take one particle from the lower left and one from the upper right corner to form a doublon in the middle (white) region, see Fig.~\ref{fig:Krylovsectorscaling}c. By choosing an even number smaller than $\frac{L^2}{12}$ of particles from both corners, it is guaranteed that the energy $E_{\Delta+U}$, see Eq.~\eqref{eq:equipotential}, can be conserved when forming the frozen pattern. In order to derive a lower bound for the scaling with system size, we count the number of possibilities to choose $n$ particles from the lower left corner to form the frozen pattern. This is given by:
\begin{align}
    N_{\text{frozen}}>\lim \limits_{L \to \infty}\sum_{n=0}^{L^2/12} \genfrac(){0pt}{}{L^2/6}{n} &> \lim \limits_{L \to \infty}\genfrac(){0pt}{}{L^2/6}{L^2/12}\\
    &= \lim \limits_{L \to \infty}\frac{(L^2/6)!}{(L^2/12)!}\\
    &=\sqrt{\frac{12}{\pi L^2}}2^{\frac{L^2}{6}}.
\end{align}
We therefore have shown, that the number of frozen states grows exponentially with system size. Next we want to generalize this argument to show, that the number of sectors with mobility show the same scaling. Mobility can be induced by e.g. placing two bosons in the middle stripe separately on empty sites instead of forming a doublon. The same argument as for the frozen patterns can then be performed.
Thus, the number of frozen, mobile and total number of Krylov sectors grows exponentially with system size.

\section{Cellular Automaton Approach}
\label{sec:Cellular_Automaton}

\begin{figure}
\centering
\tikz\path
  node[inner sep=0pt] (cellularautomat) 
    {\includegraphics[width=0.7\linewidth]{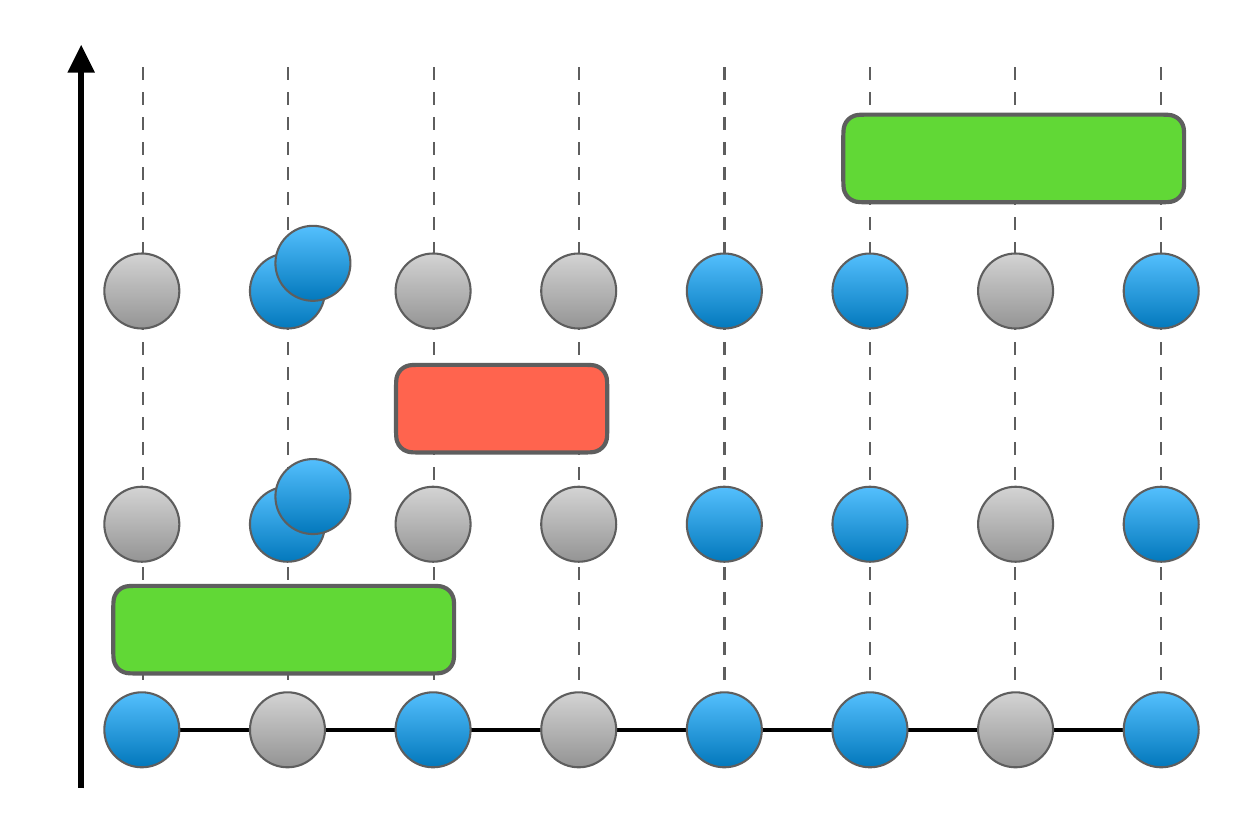}}
    (-2.8,0) node [rotate=90] {time};
    \caption{\textbf{Cellular Automaton.} Illustration of an example of an automaton circuit. For one update, first one hopping process on a site is chosen. If the process is applicable (green box) the update is performed.}
    \label{fig:Cell_auto}
\end{figure}

Studying the full quantum time evolution of the effective Hamiltonian for
large system sizes is impossible. However, we can make
progress by considering a classical automaton time evolution, which mimics the action of the Hamiltonian.
The discrete-time evolution consists of local updates,
which are given by the hopping terms of the effective
Hamiltonian. Thus they respect the same conservation
laws and map product states to product states. Time
evolution is then given by a classical simulable circuit, see Fig.~\ref{fig:Cell_auto}.
For each update, one hopping process on a random site
is chosen. If the process is applicable, then the configuration is updated, else a new update is done. The average is
performed over several runs.

\end{document}